\documentclass[aps,twocolumn,showpacs]{revtex4} 
\usepackage{graphicx}

\begin{document} 

\newcommand{\beq}{\begin{equation}} 
\newcommand{\eeq}{\end{equation}} 
\newcommand{\ben}{\begin{eqnarray}} 
\newcommand{\een}{\end{eqnarray}} 
\newcommand{\bea}{\begin{array}} 
\newcommand{\eea}{\end{array}} 
\newcommand{\om}{(\omega )} 
\newcommand{\bef}{\begin{figure}} 
\newcommand{\eef}{\end{figure}} 
\newcommand{\leg}[1]{\caption{\protect\rm{\protect\footnotesize{#1}}}}
\newcommand{\ew}[1]{\langle{#1}\rangle} 
\newcommand{\be}[1]{\mid\!{#1}\!\mid} 
\newcommand{\no}{\nonumber} 
\newcommand{\etal}{{\em et~al }} 
\newcommand{\geff}{g_{\mbox{\it{\scriptsize{eff}}}}} 
\newcommand{\da}[1]{{#1}^\dagger} 
\newcommand{\cf}{{\it cf.\/}\ } 
\newcommand{\ie}{{\it i.e.\/}\ }

\title{\center{Dynamical fidelity of a solid-state quantum computation }}

\author{G.P. Berman$^{[a]}$, F. Borgonovi$^{[b,c]}$, G. Celardo$^{[b,d]}$,
F.M. Izrailev$^{[e]}$, and D.I.Kamenev$^{[a]}$ } 
\affiliation{$^{[a]}$Theoretical Division and CNLS, Los Alamos National
Laboratory, Los Alamos, New Mexico 87545\\ $^{[b]}$Dipartimento di
Matematica e Fisica, Universit\`a Cattolica, via Musei 41, 25121
Brescia, Italy  \\ $^{[c]}$ I.N.F.M., Unit\'a di Brescia and
I.N.F.N., Sezione di Pavia, Italy\\ $^{[d]}$ I.N.F.M., Unit\`a di
Milano, Italy\\ $^{[e]}$Instituto de F\'isica, Universidad
Aut\'onoma de Puebla, Apdo. Postal J-48, Puebla 72570, Mexico\\  }
%
%
\begin{abstract}
In this paper we analyze the dynamics in
a spin-model of quantum computer. Main attention is paid to the
dynamical fidelity (associated with dynamical errors) of an algorithm
 that allows to create an entangled state for remote qubits.
We show that in the regime of selective resonant excitations of
qubits there is no any danger of quantum chaos. Moreover, in this
regime a modified perturbation theory gives an adequate
description of the dynamics of the system. Our approach allows to
explicitly describe all peculiarities of the evolution of the
system under time-dependent pulses corresponding to a quantum
protocol. Specifically, we analyze, both analytically and
numerically, how the fidelity decreases in dependence on the model
parameters.
\end{abstract}
\maketitle
\section{Introduction}

Many suggestions for an experimental realization of quantum
computers are related to two-level systems ({\it qubits}). One of
the serious problems in this field is a destructive influence of
different kinds of errors that may be dangerous for the stability
of quantum computation protocols. In the first line, one should
refer to finite temperature effects and interaction of qubits with
an environment \cite{CLSZ95}. However, even in the case when these
features can be neglected, errors can be generated by the dynamics
itself. This ``dynamical noise'' can not be avoided since the
interaction between qubits and with external fields are both
necessary for the implementation of any quantum protocol. On the
other hand, the inter--qubit interaction may cause the errors.
Therefore, it is important to know to what extent the interaction
effects may be dangerous for quantum computation.

As is known from the theory of interacting particles, a two-body
interaction between particles may result in the onset of chaos and
thermalisation, even if the system under consideration consists of
a relatively small number of particles (see, for example, the
reviews \cite{zele,GMW99,kota} and references therein). In
application to quantum computers, quantum chaos may play a
destructive role since it increases the system sensitivity  to
external perturbations. Simple estimates obtained for systems of
$L$ interacting spins show that with an increase of $L$ the chaos
border decreases, and even a small interaction between spins may
result in chaotic properties of eigenstates and spectrum
statistics. On this ground, it was claimed \cite{dima} that
quantum chaos for a large number of qubits can not be avoided, and
the idea of a quantum computation meets serious problems.

On the other hand, recent studies \cite{our} of a realistic
1/2-spin model of a quantum computer show that, in the presence of
a magnetic field gradient, the chaos border is independent on $L$,
and that quantum chaos arises in extreme situations only, which
are not interesting from the practical viewpoint. One should
stress that a non-zero gradient magnetic field is necessary in the
model \cite{our} for a selective excitation of different qubits
under time-dependent electromagnetic pulses providing a specific
quantum protocol.

Another point that should be mentioned in the context of quantum
chaos is that typical statements about chaos refer to stationary
eigenstates and spectrum statistics. However, quantum computation
is essentially a time-dependent problem. Moreover, the time of
computation is restricted by the length of a quantum protocol.
Therefore, even if stationary Hamiltonians for single pulses
reveal chaotic properties, it is still not clear to what extent
stationary chaos influences the evolution of a system subjected to
a finite number of pulses.

In contrast with our previous studies \cite{our}, in this paper we
investigate a time evolution of a 1/2-spin quantum computer system
subjected to a series of pulses. Specifically, we consider a
quantum protocol that allows to create an entangled state for
remote qubits. For this, we explore the model in the so-called
{\it selective} regime, using both analytical and numerical
approaches. Our analytical treatment shows that in this regime
there is no any fingerprint of quantum chaos. Moreover, we show
that a kind of perturbation approach provides a complete description
of the evolution of our system.

We concentrate our efforts on the introduced quantity ({\it
dynamical fidelity}). This quantity characterizes the performance
of quantum computation associated with the {\it dynamical} errors.
Dynamical fidelity differs from the fidelity which is associated
with the external errors, and which is widely used nowadays in
different applications to quantum computation and quantum chaos,
see for instance \cite{allqc}. Our study demonstrates an excellent
agreement of analytical predictions with numerical data.

The structure of the paper is as follows. In  Section II we
discuss our model and specify the region of parameters for which
our study is performed. In Section III we explore a possibility of
quantum chaos in the selective regime, and analytically show that
chaos can not occur in this case. We provide all details
concerning the quantum protocol in Section IV, and demonstrate how
perturbation theory can be applied to obtain an adequate
description of the fidelity in dependence on the system
parameters. Here, we also present numerical data and compare them
with the predictions based on the perturbative approach. The last
Section V summarizes our results.


\section{Spin model of a quantum computer}

Our model is represented by a one-dimensional chain of $L$
identical $1/2$-spins placed in an external magnetic field.
It was first proposed in \cite{ber1} (see also
\cite{BDMT98,ber1.1,ber1.2}) as a simple model for solid-state
quantum computation. Some physical constraints are necessary in
order to let it operate in a quantum computer regime. To provide a
selective resonant excitation of spins, we assume that the time
independent part $B^z = B^z(x)$ of a magnetic field is 
non-uniform along the spin chain. The non-zero gradient of the
magnetic field provides different Larmor frequencies for different
spins. The angle $\theta$ between the direction of the chain and
the $z$-axis satisfies the condition, $\cos \theta = 1/\sqrt{3}$.
In this case the dipole-dipole interaction is suppressed, and the
main interaction between nuclear spins is due to the Ising-like
interaction mediated by the chemical bonds, as in a liquid state
NMR quantum computation \cite{CLSZ95}.

In order to realize quantum gates and implement operations, it is
necessary to apply selective pulses to single spins. The latter
can be distinguished, for instance, by imposing constant gradient
magnetic field that results in the Larmor frequencies $\omega_k =
\gamma_n B^z (x_k) =\omega_0+a k$, where $\gamma_n$ is the spin
gyromagnetic ratio and $a k=x_k$
and $x_k$ is the position of the $k$-th spin.
If the distance between the
neighboring nuclear spins is $\Delta x=0.2$ nm, and the frequency
difference between them is $\Delta f=a/2\pi=1$ kHz, then the
corresponding gradient of the magnetic field can be estimated as
follows, $|dB^z/dx|=\Delta f/(\gamma_n/2\pi)\Delta x\approx 1.2
\times 10^4$ T/m. Here we used the gyromagnetic ratio for a
proton, $\gamma_n/2\pi\approx 4.3\times 10^7$ Hz/T. Such a
magnetic field gradient is experimentally achievable, see, for
example, \cite{sidles,suter}.

In our model the spin chain is also subjected to a transversal
circular polarized magnetic field. Thus, the expression for the
total magnetic field has the form \cite{our,ber1.1,ber1.2},
\begin{equation}
{\vec B}(t)=[b^p_\perp\cos(\nu_p t+\varphi_p),
-b^p_\perp\sin(\nu_p t+\varphi_p), B^z(x)].
\end{equation}

As is mentioned above, here $B^z(x)$ is the constant magnetic
field oriented in the positive $x$-direction, with a positive
$x$-gradient (therefore, $a>0$ in the expression for the Larmor
frequencies). In the above expression, $b^p_\perp$, $\nu_p$, and
$\varphi_p$ are the amplitudes, frequencies and phases of a
circular polarized magnetic field, respectively. The latter is
given by the sum of $p=1,...,P$ rectangular time-dependent pulses
of length $t_{p+1}-t_p$, rotating in the $(x,y)-$ plane and
providing a quantum computer protocol.

Thus, the quantum Hamiltonian of our system has the form,
\begin{equation}
\begin{array}{ll}
{\cal H}= -\sum\limits^{L-1}_{k=0} \left[\omega_kI^z_k+2
\sum\limits_{n > k}J_{k,n} I^z_k I^z_n \right]-
\\ ~~~~~~~~~\\
{{1}\over{2}}\sum\limits_{p=1}^{P}\Theta_p(t)\Omega_p
\sum\limits_{k=0}^{L-1} \Bigg(e^{-i\nu_p
t-i\varphi_p}I^-_k+ e^{i\nu_p t+i\varphi_p}I^+_k\Bigg),
\label{ham00}
\end{array}
\end{equation}
where the ``pulse function" $\Theta_p(t)$ equals $1$ only during
the $p$-th pulse, for $t_p< t\le t_{p+1}$, otherwise, it is zero.
The quantities $J_{k,n}$ stand for the Ising interaction between
two qubits , $\omega_k$ are the frequencies of  spin precession
in the $B^z-$magnetic field, $\Omega_p$ is the Rabi frequency of
the $p$-th pulse, $I_k^{x,y,z} = (1/2) \sigma_k^{x,y,z}$ with $
\sigma_k^{x,y,z}$ as the Pauli matrices, and
$I_k^{\pm}=I^x_k \pm iI^y_k$.

For a specific $p$-th pulse, it is convenient to represent the
Hamiltonian (\ref{ham00}) in the coordinate system that rotates
with the frequency $\nu_p$. Therefore, for the time $t_p<t\le
t_{p+1}$ of the $p$-th pulse our model can be reduced to the {\it
stationary} Hamiltonian,
\begin{equation}
\begin{array}{ll}
{\cal H}^{(p)}=-\sum\limits_{k=0}^{L-1} (\xi_k I^z_k+
\alpha I^x_k-\beta I^y_k)-
2  \sum\limits_{n>k}^{}J_{k,n}I^z_k I^z_n,
\label{ham}
\end{array}
\end{equation}
where $\xi_k=(\omega_k-\nu_p)$, $\alpha=\Omega_p \cos \varphi_p$,
and $\beta = \Omega_p \sin \varphi_p$.

We start our considerations with the simplified case of the
Hamiltonian (3) for a single pulse, by choosing $\varphi_p=0$. We
also assume a constant interaction between nearest qubits only
($J_{k,n}=J\delta_{k,k+1}$), and we put $\Omega_p=\Omega$. Then
the Hamiltonian (3) takes the form,
\begin{equation}
{\cal H}^{(p)}=\sum_{k=0}^{L-1}
\Big [-\xi_k  I^z_k- 2J I^z_k I^z_{k+1} \Big]
- \Omega \sum_{k=0}^{L-1} I^x_k \equiv H_{0}+V.
\label{ham0}
\end{equation}

In the $z$-representation the Hamiltonian matrix of size $2^L$ is
diagonal for $\Omega=0$. For $\Omega\not=0$, non-zero off-diagonal
matrix elements are simply $H_{kn}=H_{nk}=- \Omega/2$ with $n
\neq k$. The matrix is very sparse, and it has a specific
structure in the basis reordered according to an increase of the
number $s$. The latter is written in the binary representation,
$s=i_{L-1},i_{L-2},...,i_{0}$ (with $i_s=0$ or $1$, depending on
whether the single-particle state of the $i-$th qubit is the
ground state or the excited one). The parameter $\Omega$ thus is
responsible for a non-diagonal coupling, and we hereafter define
it as a ``perturbation''.

In our previous studies \cite{our} we have analyzed the so-called
{\it non-selective} regime which is defined by the conditions,
$\Omega_p\gg \delta\omega_k \gg J$. This inequality provides the
simplest way to prepare a homogeneous superposition of $2^L$
states needed for the implementation of both Shor and Grover
algorithms. Our analytical and numerical treatment of the model (2) in
this regime has shown that a constant gradient magnetic field (with
non-zero value of $a$) strongly reduces the effects of quantum chaos.
Namely, the chaos border turns out to be independent on the number
$L$ of qubits. As a result, for non-selective excitation quantum
chaos can be practically neglected (see details in
\cite{our}).

Below we consider another important regime called {\it selective
excitation}. In this regime each pulse acts selectively on a
chosen qubit, resulting in a resonant transition. During the
quantum protocol, many such resonant transitions take place for
different $p$ pulses, with different values of $\nu_p=\omega_k$.
The region of parameters for the selective excitation is specified
by the following conditions \cite{ber1.1},
\begin{equation}
\Omega_p \ll J_{k,n} \ll a \ll\omega_k.
\label{con2}
\end{equation}
The meaning of these conditions will be discussed in next
Sections.

\section{Absence of quantum chaos in the selective excitation regime}

Here, we consider the properties of the stationary Hamiltonian
(\ref{ham0}) in the regime of selective excitation. In order to
estimate the critical value of the interaction $J$, above which
one can expect chaotic properties of eigenstates, one needs to
compare the typical value of the off-diagonal matrix elements
(which is $\Omega/2$) with the mean energy spacing $\delta_f$ for
unperturbed many-body states that are directly coupled by these
matrix elements. Therefore, the condition for the onset of chaos
has the form,
\begin{equation}
\displaystyle\frac{\Omega}{2}  > \delta_f \approx
 \frac{(\Delta E)_f}{M_f}.
\label{esti}
\end{equation}
Here $(\Delta E)_f$ is the maximal difference between the energies
$E_0^{(2)}$ and $E_0^{(1)}$  corresponding to a specific many-body
state $|1\rangle$ , and all other states $|2\rangle$ of $H_0$,
that have non-zero couplings $\langle 1|V|2\rangle$.
Correspondingly, $M_f$ is the number of many-body states
$|2\rangle$ coupled by $V$ to the state $|1\rangle$. A further
average over all states $|1\rangle$ should be then performed.

In fact, such a comparison (\ref{esti}) is just the perturbation
theory in the case of  two-body interaction. Strictly speaking,
the above condition in a strong sense ($\Omega \gg \delta_f$)
means that exact eigenstates consist of many unperturbed ($V=0$)
states. Typically, the components of such compound states can be
treated as uncorrelated entries, thus resulting in a chaotic
structure of excited many-body states. However, one should note
that in specific cases when the total Hamiltonian is integrable
(or quasi-integrable), the components of excited states have
strong correlations and can not be considered as chaotic, although
the number of components with large amplitudes can be extremely
large (see details in \cite{our}).

It is relatively easy to estimate $M_f$ in the regime of
selective excitation. Let us consider an eigenstate of $H_0$,
 $|1,0,0,0,1,0,...,0,0,1,0\rangle$, as a collection of $0$'s and
$1$'s that correspond to $-1/2$ and $1/2$-spin values. Since the
perturbation $V$ is a sum of $L$ terms, each of them flipping one
single spin, one gets $M_f \sim L$.

In order to estimate $(\Delta E)_f$, let us first consider the
action of $V$ on the $k$-th spin, and for each spin compute the
relative energy difference between the final and the initial
energy. One can find that if the $k$-th spin has the value $1/2$,
there are four possible configurations of neighbor spins coupled
by the perturbation,
\begin{equation}
\begin{array}{lll}
|...0,1,0...\rangle &  \to |...0,0,0...\rangle, \\
|...1,1,1...\rangle &  \to |...1,0,1...\rangle, \\
|...1,1,0...\rangle &  \to |...1,0,0...\rangle, \\
|...0,1,1...\rangle &  \to |...0,0,1...\rangle. \\
\end{array}
\label{tru}
\end{equation}
If the $k$-th spin has the value $-1/2$, there are also four
possible different arrangements,
\begin{equation}
\begin{array}{lll}
|...0,0,0...\rangle &  \to |...0,1,0...\rangle, \\
|...1,0,1...\rangle &  \to |...1,1,1...\rangle, \\
|...1,0,0...\rangle &  \to |...1,1,0...\rangle, \\
|...0,0,1...\rangle &  \to |...0,1,1...\rangle, \\
\end{array}
\end{equation}
which are the inverse  transitions of (\ref{tru}).
Correspondingly, the energy changes are determined by the
relation,
\begin{equation}
\label{cond1}
 |E_0^{(f)}-E_0^{(i)} |=  |\xi_k \pm 2J |,|\xi_k|,\qquad k=1,...,L-2.
\end{equation}
The analysis for the border spins can be performed in a similar
way, and one gets four possible configurations, with the following
energy changes,
\begin{equation}
\label{cond2}
 |E_0^{(f)}-E_0^{(i)} |= |\xi_k \pm J |,\qquad k=0,L-1.
\end{equation}
Summarizing the above findings, and setting for instance
$\nu_p = \omega_0$,  one can conclude that $(\Delta
E)_f$ can be estimated as follows,
\begin{equation}
(\Delta E)_f = Max ( |E_0^{(f)}-E_0^{(i)} |) = \omega_{L-1}-\omega_0 +J.
\label{dele}
\end{equation}

As a result, the condition for the onset of quantum chaos can be
written in the form,
\begin{equation}
\displaystyle\frac{\Omega}{2} >
\frac{(\Delta E)_f}{M_f}= \frac{\omega_{L-1}-\omega_0+J}{L}
= \frac{a(L-1)+J}{L},
\label{cro}
\end{equation}
or
\begin{equation}
\displaystyle \Omega > \Omega_{cr} \simeq a  + \frac{J}{L}.
\label{cri}
\end{equation}

However, this critical value is outside the range of parameters
required to be in the selective excitation regime $\Omega <
a$ (see inequality (6)). Thus, we can conclude that quantum
chaos for stationary states
can not appear in the selective excitation regime. Note that the
analysis is done for a single pulse of a time-dependent
perturbation.

\section{Fidelity of a quantum protocol}

Analytical results obtained above, show that during a single
electromagnetic pulse the system can be described by  perturbation
theory. Indeed, if matrix elements of perturbation are smaller
than the energy spacing between directly coupled many-body states,
exact eigenstates can be obtained by perturbation theory. Thus, one can
expect that for a series of time-depended pulses the evolution of
the systems can  be also treated making use of a kind of
perturbative approach.

In what follows, we study the dynamics of the system by applying
 a specific set of pulses (quantum protocol) in order to create an
entangled state for remote qubits (with $k=0$ and $k=L-1$)
starting from the ground state,
$|\psi_0\rangle=|0_{L-1},...,0_1,0_0\rangle$ (we omit the
subscripts below). Our main interest is in estimating the errors
that appear due to unwanted excitations of qubits. We show that
these errors can be well understood and estimated on the basis of
the perturbation theory developed for our time-dependent
Hamiltonian (\ref{ham00}), in the parameter range where the
protocol holds.

\subsection{Selective excitation regime and perturbation theory}

Any protocol is a sequence of unitary transformations applied to
some initial state in order to obtain a final state,
$|\psi^i\rangle$. In this model of quantum computer the protocol
is realized by applying a number of specific rf-pulses, so that we
get a state  $|\psi^r\rangle$ which is, in principle, different
from the ideal state $|\psi^i\rangle$. The difference between the
final state $|\psi^r\rangle$ and the ideal state $|\psi^i\rangle$
can be characterized by a {\it dynamical fidelity},
\begin{equation}
\label{fidel}
F=|\langle \psi^i|\psi^r\rangle |^2.
\end{equation}
Note that, in our case the dynamical fidelity $F$ does not
explicitly depend on a perturbation parameter added in the
Hamiltonian (2) in order to get a distorted evolution, as is
typically assumed in the study of quantum chaos. Indeed, the final
state is determined by the total Hamiltonian (2),
\begin{equation}
|\psi^r\rangle = {\hat U}(T) |\psi_0 \rangle
\equiv \prod_{p=1}^P  \hat{T} e^{-i \int_{t_{p-1}}^{t_p}
H (t)dt} |\psi_0
\rangle,
\end{equation}
where $T= t_p$ is the total time to entangle spins, $\hat U(T)$ is
the unitary operator given by the sequence of pulses in the
protocol, and $\hat{T}-$ is the usual time-ordered product.
Therefore, it is not possible to identify a single perturbation
parameter which is responsible for a ``wrong'' evolution of the
system.

The selective excitation regime is characterized by the action of
pulses that are resonant with a transition between two energy
states which differ for the state (up or down) of one  spin only.
A close inspection of the time independent Hamiltonian
(\ref{ham0}), defines the region of parameters where the selective
excitation of single spins can be performed.

Diagonal elements of Hamiltonian (\ref{ham0}) are given by the
eigenvalues $E_0^{(i)}$ of $H_0$, while non-zero off-diagonal
elements are constant and equal to $- \Omega/2$. In order to have
a resonant transition between two energy states, their energy
difference $\Delta$ has to be zero. On the other side, for each
state no more than one resonant transition should be allowed. So,
we require that the energy differences given by Eqs.
(\ref{cond1},\ref{cond2}) are  different from zero (apart
from the wanted transition). This leads to the following set of
equalities (``fake transitions''),
\begin{equation}
\begin{array}{lll}
J &= a \frac{k}{4} \quad {\rm when}  \quad k=1,...,L-3,\\
J &= a \frac{k}{2} \quad {\rm when}  \quad k=1,...,L-3,\\
J &= a k\quad {\rm when}  \quad k=1,...,L-2,\\
J &= a \frac {k}{3}  \quad {\rm when}  \quad k=1,...,L-2,\\
\label{gein}
\end{array}
\end{equation}

From Eqs.(\ref{gein}) it is easy to see that first ``fake''
transition  appears for $J_1= a/4$, the second for $J_2=a/2$ and
so on up to the last one for $J_f=a(L-2)$. All these resonances
can be avoided if we choose $a \gg 4 J$. Indeed, it is not enough
to chose simply $a>4J$ since resonances have finite widths.

Transitions can be defined according to their energy difference $
\Delta$,
\begin{itemize}
\item[1)] {\it resonant transitions}, $\Delta =0$;
\item[2)] {\it near--resonant transitions}, $\Delta \sim J$;
\item[3)] {\it non--resonant transitions}, $\Delta \sim a $.
\end{itemize}

For $a \gg 4 J$, each state can undergo one
resonant or near-resonant transition only, and many non-resonant
ones. The latter can be neglected if we choose $a \gg  \Omega$.
Under these conditions we can form couples of states, connected by
resonant or near-resonant transitions, and we can rearrange the
Hamiltonian matrix (\ref{ham0}) by $2 \times 2$ block matrices
representing all resonant and near-resonant transitions. This
allow us to describe the dynamical evolution of the system as a
two-state problem.

Using this procedure, the entire sequence of pulses can be
evaluated. Note that special attention has to be paid to an
additional phase shift that arises between any two pulses, due to
the change of frame. We remind that the transformation  between
the rotating and the laboratory frame is given by the expression,
\begin{equation}
\displaystyle
|\psi(t) \rangle_{Lab}= e^{i\nu_{p}t \sum_k I^z_k }|\psi(t) \rangle_{Rot}
\end{equation}

Indeed, let us consider an initial basis state $|m\rangle$ at time
$t=0$, and find the probability for a resonant ($\Delta =0$) or
near-resonant ($\Delta \sim J$) transition to the state
$|p\rangle$  with the energy difference $E_p-E_m$. Here $E_p$ and
$E_m$ are the eigenenergies of the time-independent part of the
Hamiltonian (\ref{ham00}), written in the laboratory frame.

Setting,
\begin{equation}
\psi(t) = \sum_n c_n(t) e^{-iE_n t} |n\rangle,
\label{ppp}
\end{equation}
and $c_p(0)=0$, the application of a pulse for a time $\tau$ gets:
\begin{equation}
\begin{array}{lll}
c_m(\tau) &=c_m(0)[\cos(\frac{\lambda \tau}{2})+
i \ \frac{\Delta}{\lambda} \ \sin(\frac{\lambda \tau}{2})]
e^{-i\frac{\Delta\tau}{2}-iE_m\tau},\\
c_p(\tau) &=
c_m(0)[i \ \frac{\Omega}{\lambda} \ \sin(\frac{\lambda \tau}{2})]
e^{i\frac{\Delta\tau}{2}-iE_p\tau},\\
\label{raf}
\end{array}
\end{equation}
where  $\lambda=\sqrt{\Omega^2+\Delta^2}$. Note that
Eqs.(\ref{raf}) refer to the laboratory frame.

As we can see, the parameter $\epsilon$ determined as
\begin{equation}
\epsilon = \frac{\Omega^2}{\Omega^2+\Delta^2}
\sin\left(\frac{\tau}{2} \sqrt{\Omega^2 + \Delta^2}\right)
\label{eps}
\end{equation}
characterizes the probability of resonant and near-resonant
transitions. In particular, the probability of unwanted
near-resonant transitions goes like $\epsilon $, and it can be
reduced by assuming $J \gg \Omega$. Combining all the above
expression, we get the condition (\ref{con2}).

Correspondingly, the probability for a non-resonant transition
(neglecting terms of the order $1/L$, and assuming $a \gg \Omega$
) is given by the parameter $\eta$ \cite{ber1.2},
\begin{equation}
\eta = \frac{\Omega^2}{4 a^2} .
\label{diom}
\end{equation}

We would like to stress that even if the ideal state has been
constructed taking into account resonant transitions only, our
dynamical fidelity is a measure of dynamical errors that are due
to near and non-resonant transitions.

Let us now briefly discuss the perturbative approach that is based
on recent studies published in Ref.\cite{ber1.2}. The main idea is
that for each $p$-th pulse the unperturbed basis can be rearranged
in such a way that the Hamiltonian matrix is represented by
$2\times 2$ block matrices, as described above. This is what we
call {\it unperturbed} Hamiltonian for a specific $p$-th pulse.
One should note that this {\it unperturbed} Hamiltonian is
$\Omega$-dependent. Let us now define by ${\cal V}$ the
$\Omega$-dependent part which is responsible for non-resonant
transition and not described by the $2\times 2$ block matrices.
Then it is easy to obtain the unperturbed eigenstates, $|\psi_q^0
\rangle$, and the unperturbed eigenvalues, $\epsilon_q^0$, by
diagonalizing each of the $2\times 2$ blocks independently.

After this step, one can compute the {\it perturbed} eigenstates by
taking into account first order terms only,
\begin{equation}
|\psi_q \rangle =|\psi_q^0 \rangle +\sum_{q' \ne q} \frac{ \langle
\psi_q^0|{\cal V}|\psi_{q'}^0 \rangle}
{\epsilon_q^0-\epsilon_{q'}^0}|\psi_{q'}^0\rangle.
\label{pt1}
\end{equation}

Note that this perturbative approach is supposed to be valid when
Eqs.(\ref{gein}) are not satisfied, and when the errors due to
near-resonant transitions are much larger than the errors due to
non-resonant ones, $\epsilon \gg \eta$.

\subsection{Quantum protocol}

Let us briefly sketch the algorithm and the particular protocol
which was developed in Ref. \cite{ber1.1}. Starting from the
ground state $|\psi_0\rangle = |0...0\rangle$ and applying a
number of specific pulses, we would like to generate the following
entangled state,
\begin{equation}
\label{eq1}
|\psi^i\rangle=\frac{1}{ \sqrt{2}} \left( |0...0\rangle+ |10....01
\rangle \right).
\end{equation}
\noindent
This algorithm could serve, for instance, as the first step for a
more general teleportation protocol, and for an implementation of
conditional quantum logic operations.

The algorithm can be realized in the following way (for details
see Ref. \cite{ber1.1}),
\begin{equation}
\begin{array}{lll}
|0,...,0 \rangle &\rightarrow (|0,...,0 \rangle +|1,0,..,0
\rangle) \\ &\rightarrow (|0,...,0 \rangle +|1,1,0,..,0\rangle)\\
&\rightarrow (|0,...,0 \rangle +|1,1,1,0,..,0\rangle)\\
&\rightarrow (|0,...,0 \rangle +|1,0,1,0,..,0\rangle)\\
&\rightarrow \ldots \rightarrow (|0,...,0\rangle +|1,0,..,1
\rangle ).
\label{prot}
\end{array}
\end{equation}
Physically, the above algorithm can be done by applying suitable
{\it rf}-pulses that are resonant to the desired transitions. The
latter are originated from induced Rabi oscillations between the
resonant states.

To flip  the $k$-th spin we have to choose the frequency $\nu$ of
the {\it rf}-pulse according to the relation $\nu_p =E_1-E_2$,
where $|1\rangle$ and $|2\rangle$ are the states involved in the
transition and $E_1$ and $E_2$ are the eigenenergies of the
time-independent part of Hamiltonian (\ref{ham00}). For instance,
for the first pulse we have to set,
$\nu_1=|E_{|1,0,...,0\rangle}-E_{|0,...,0\rangle}|$, and we have
to apply it for a time $t_1=\pi/2\Omega$ to get equal
superposition of the states involved in the transition. For other
pulses we require that the first state ($|0,...,0\rangle$) remains
the same (apart from an additional phase), while the second state
flips the $k-$th spin. This amounts to say that the probability of
unwanted states is due to non-resonant transitions of both states,
and to near-resonant ones only of the first state in the r.h.s. of
Eq.(\ref{prot}). Specifically, the state $|0,...,0\rangle$
undergoes near-resonant transitions with $\Delta=2J$ for each
pulse, except the first one which is resonant, and the forth for
which $\Delta=4J$. One should also take into account that at each
pulse the state $|0,...,0\rangle$ get an additional phase, see
eqs. (\ref{raf}). We took them into account in the definition of
the ideal state.

Since in the selective excitation regime we have $\epsilon \gg
\nu$, contributions from near-resonant transitions are much larger
than the one due to non-resonant transitions. Our algorithm
consists of $2L-2$ separate pulses, therefore, some modifications
are necessary in order to be able to control small unwanted
probability. For the product of probabilities this requires to
impose, $2L\epsilon < 1$ and $2L\eta < 1$, that can be written in
the form,
\begin{equation}
\frac{\Omega}{J} \ll \sqrt{ \frac{2}{L}},\,\,\,\,\,\,\,\,\,\,\,\,
\frac{\Omega}{a} \ll \sqrt{ \frac{2}{L}}
\label{coal}
\end{equation}
where $L \gg 1$.

Before discussing our numerical results we would like to stress
that in contrast to what is mainly considered in the literature,
the time for our dynamical fidelity is not an independent
variable. Indeed, the length of the protocol is determined by the
total number of qubits, $L$. Specifically, $2L-2$ pulses are
necessary in order to create the entangled state, so that  the
protocol time $T$ is proportional to the number of qubits.

\subsection{Dynamical fidelity: theory and numerical data}

Quite unexpectedly, the dynamical fidelity (14) increases with an
increase of the Ising coupling $J$, as soon as $J \ll a/4$. This
is due to the fact that the probability of unwanted near-resonant
transitions is proportional to $\epsilon \sim (\Omega/J)^2$ (see
Eq.(\ref{eps}) where $\Delta\sim J$ for near-resonant transitions
). The larger is $J$, the smaller is the probability of
non-resonant transitions, therefore, the dynamical fidelity $F$
increases with an increase of $J$.

\begin{figure}
\includegraphics[scale=0.46]{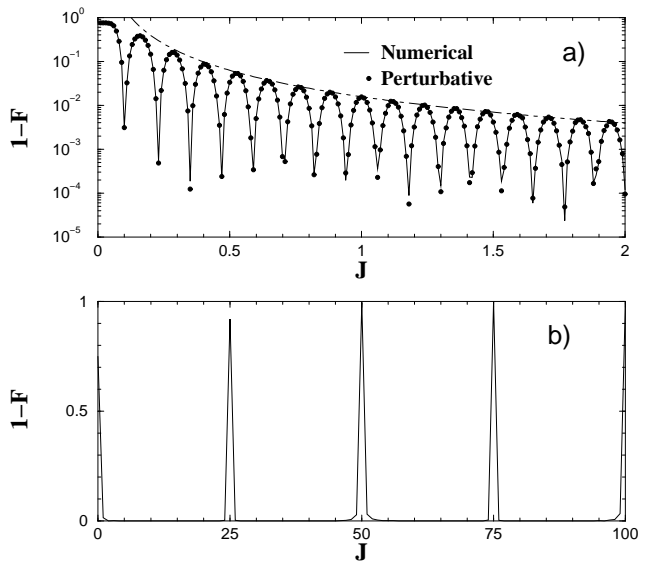}
\caption{The dependence $1-F$ is shown as a function of the
Ising coupling $J$ for $L=6$ spins, $\Omega=0.118$ and $a=100$.
Full line represents the numerical data for the dynamical fidelity
$F$ defined by Eq.(\ref{fidel}), and obtained from direct
numerical computation of the system evolution. (a) Full circles
stand for perturbative calculations, and full curve corresponds to
numerical results. (b) The same numerical results as in (a), but
for a larger range of $J$. The theoretical expression as given by
Eq.(\ref{fidim1}) is also shown in (a).}
\label{fidj}
\end{figure}

In Fig.\ref{fidj} we show how the dynamical fidelity (\ref{fidel})
depends on the inter-qubit interaction $J$. For convenience, the
function $1-F$ is shown here and below, instead of $F$. Numerical
data have been obtained in two different ways. Full curve
corresponds to exact computation of the time-dependent Hamiltonian
(\ref{ham00}). Data in Fig.\ref{fidj}a are compared with those
obtained from the perturbative approach explained above.

Apart from very strong peaks (see Fig.\ref{fidj}b) for which the
dynamical fidelity vanishes, one can say that the global tendency
is an improvement of the dynamical fidelity for larger values of
$J$. However, strong oscillations occur reflecting a resonant
nature of the dynamics of our system. Perfect agreement between
perturbative results and numerical data is found for very large
variations of the interaction strength $J$.

High peaks for $1-F$, clearly seen in Fig.\ref{fidj}b, are due to
Eqs.(\ref{gein}), in these points the quantum algorithm fails.
Thus, one should avoid these situations in a quantum computation.
As for the minima in Fig.\ref{fidj}a for which the dynamical
fidelity is close to one, they occur when $\epsilon=0$, or, when
$$ J =\frac{\Omega}{2} \sqrt{4k^2-1},$$ where $k$ is an integer
number. This relation corresponds to the $2\pi k$-condition
\cite{ber1.1,ber1.2,campbell}.

\begin{figure}
\includegraphics[scale=0.46]{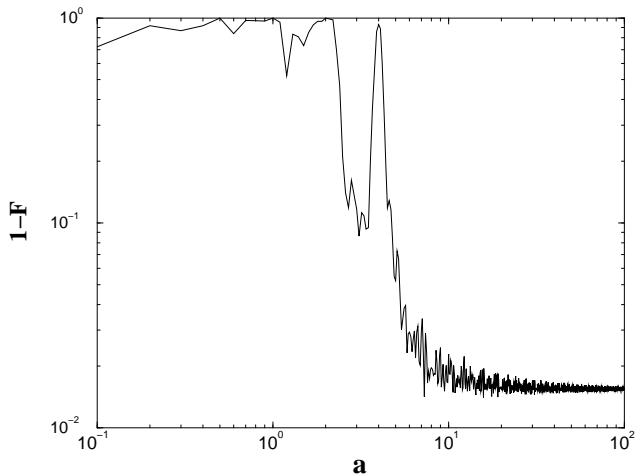}
\caption{
The dependence $1-F$ as a function of the gradient of a magnetic
field is shown for $L=6$ spins, $\Omega=0.118$ and $J=1$. As one
can see, for $a< 4 J $ the correspondence is not so good as for $a
> 4 J $. }
\label{fida}
\end{figure}

Let us now explore the dependence of the dynamical fidelity on the
parameter $a$ which is proportional to the gradient of the
external magnetic field, $a=\gamma_n\Delta x(dB^z(x)/dx)$, where
$\Delta x$ is the distance between neighboring qubits. (Below, we
shall refer to the parameter $a$ as to the magnetic field
gradient.) Numerical data for the dependence of $1-F$ on $a$ are
presented in Fig.\ref{fida}. One can see that the dynamical
fidelity is getting better for large enough values of $a$. We
already mentioned that for $a < 4J$  a problem may arise in the
protocol due to ``fake'' transitions. On the other side, in the
regime $a \gg 4J$ the dynamical fidelity reaches an asymptotic
value which depends on $J$ and $\Omega$ only, see Fig.\ref{fida}.

\begin{figure}
\includegraphics[scale=0.46]{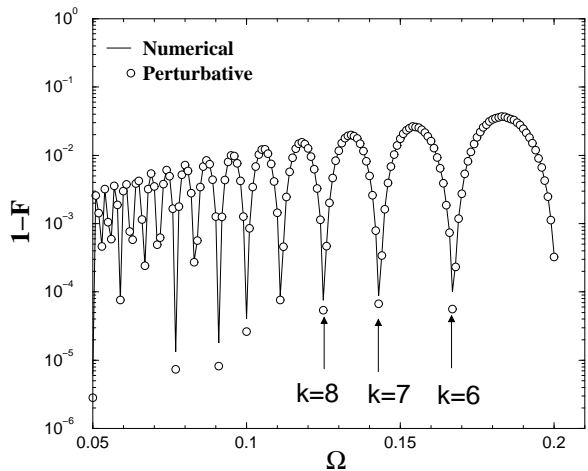}
\caption{The difference $1-F$ as a function of the Rabi frequency
$\Omega$ for $L=6$ spins, $J=1$ and $a=100$. Full curve is the
result of  direct numerical simulation, circles are obtained
from the perturbative  approach described in the text. Arrows
show few  resonant values of $\Omega$ given by Eq.(\ref{tupi}).}
\label{fidom}
\end{figure}

It is also important to understand the dependence of the dynamical
fidelity on the Rabi frequency $\Omega$. The data manifest two
specific properties demonstrated in Fig.\ref{fidom}. The first one
is a global decrease of the dynamical fidelity with an increase of
$\Omega$. The second peculiarity is due to strong oscillations
that occur for $\epsilon = 0$, namely for those $\Omega$ values
that  correspond to the $2\pi k$-conditions,
\begin{equation}
\Omega_k = \displaystyle \frac{2J }{\sqrt{4k^2 - 1}}.
\label{tupi}
\end{equation}
For these values of $\Omega$ near-resonant transitions vanish, and
non-resonant transitions remain only. Thus, the dynamical fidelity
has maxima which provides, in principle, the best condition for a
quantum computation.

Nevertheless, let us consider the values of $\Omega$ that
correspond to maxima in Fig.(\ref{fidom}). This we do in order to
make an estimate in the worst possible condition. A brief analysis
of the fidelity for specific values $\Omega=\Omega_k$ will be
sketched in the last subsection. As one can see, for values of
$\Omega$ different from $\Omega_k$, the ``average'' dynamical
fidelity increases when the Rabi frequency decreases. This is due
to the fact that the probability to generate unwanted states (due
to both non-resonant and near-resonant transitions), is
proportional to $(\Omega/\Delta)^2$. Therefore, the smaller is
$\Omega$ the more reliable the algorithm is. Note that the
agreement with the perturbative approach is excellent.

On the other hand, we cannot choose an extremely small value of
$\Omega$ since it implies a large time duration of the pulse
($\tau \sim\pi/\Omega$). Note, that the total time for a quantum
protocol should be kept well below the decoherence time (the
latter can be quite large for nuclear spins \cite{kane}). Taking
this point into account, an optimal choice is to take the largest
possible value, $\Omega = \Omega_2 = 2J/\sqrt{15}<J$, and large
enough value of $a$ (in order to significantly suppress the
non-resonant transitions).

\subsection{Fidelity: dependence on the number of qubits}

\begin{figure}
\includegraphics[scale=0.46]{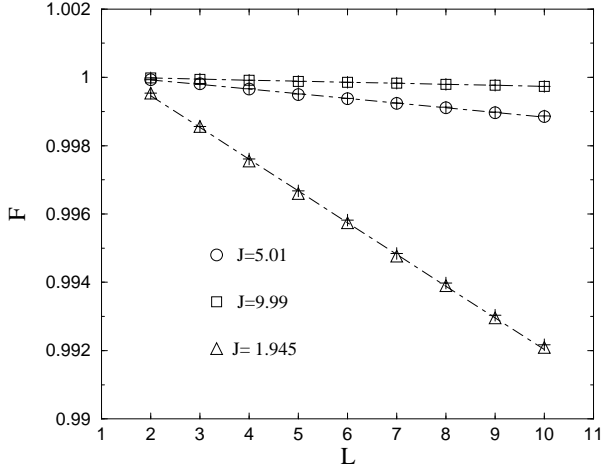}
\caption{
The dynamical fidelity as a function of the number $L$ of spins,
for different $J$ values and $\Omega=0.118$, $a=100$. Numerical
data (triangles for $J=1.945$, circles for $J=5.01$ and squares
for $J=9.99$) are compared with the results from the perturbation
theory (crosses). Also shown are the best linear fits (dot-dashed
lines).}
\label{fidn}
\end{figure}

Finally, we studied the dependence of the dynamical fidelity on
the number $L$ of spins in the chain. As was noted before, for a
chosen protocol its length is proportional to $L$. Numerical data
clearly manifest a linear decrease of the dynamical fidelity with
the number of qubits, see Fig. \ref{fidn}.

This dependence can be also understood on the basis of
perturbation theory, by neglecting the change of phases
between the ideal and the real state. This can
be done since we already considered the main change of phases in
the definition of the ideal state. Let us write,
\begin{equation}
| \psi^r \rangle = \sum_k c_k |\psi_k\rangle,
\label{svi}
\end{equation}
and assume that after $M$ pulses the components $c_0$ and $c_1$
are,
\begin{equation}
c_0 =\frac{1}{\sqrt{2}} \sqrt{1-M\epsilon};\,\,\,\,\,\,\,\, c_1
=\frac{1}{\sqrt{2}}.
\label{ansa}
\end{equation}
Here $c_0$ and $c_1$ are the complex amplitudes of two states in
Eq.(\ref{eq1}), but obtained as the result of the quantum
protocol. This means that the probability of near-resonant states
after $M$ pulses is small, $M\epsilon \ll 1$. Then we get,
\begin{equation}
F = \frac{1}{4}(2- M\epsilon +2\sqrt{1-M\epsilon})
\sim  1-M\frac{\epsilon}{2},
\label{fidim}
\end{equation}
where $\epsilon  \sim \Omega^2/4 J^2$ and $M=2L-3$ for our
specific algorithm. Combining all the above expressions, one gets,
\begin{equation}
\displaystyle F \sim
- \frac{\Omega^2}{4J^2} L + \left(1+ \frac{3\Omega^2}{8 J^2}\right),
\label{fidim1}
\end{equation}
which implies a linear decrease of the dynamical fidelity with an
increase of the number of qubits, $L$. The slope is given by the
parameter $m_{th} = -\Omega^2/4 J^2$. Of course, this
consideration is valid far from the resonant conditions
(\ref{gein}).

\begin{figure}
\includegraphics[scale=0.46]{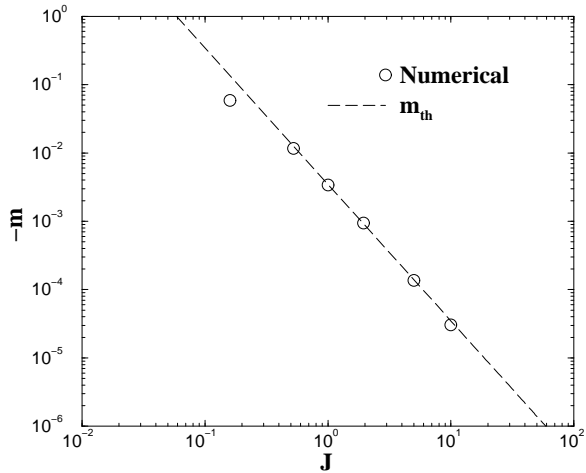}
\caption{
Comparison between theoretical and numerical linear slopes for the
fidelity, as a function of the interaction $J$.}
\label{sloppy}
\end{figure}

To compare numerical data with the above analytical estimate for
$m_{th}$, we obtained the slopes in Fig. \ref{fidn} by the best fit
to a linear dependence. These slopes versus $J$ are shown in
Fig. \ref{sloppy}, together with the theoretical result for
$m_{th}$.

As one can see, the agreement is very good except for small values
of the Ising interaction, $J\simeq \Omega$, where the probability
of near-resonant transitions becomes large.

\subsection{Optimal algorithm}

\begin{figure}
\includegraphics[scale=0.36]{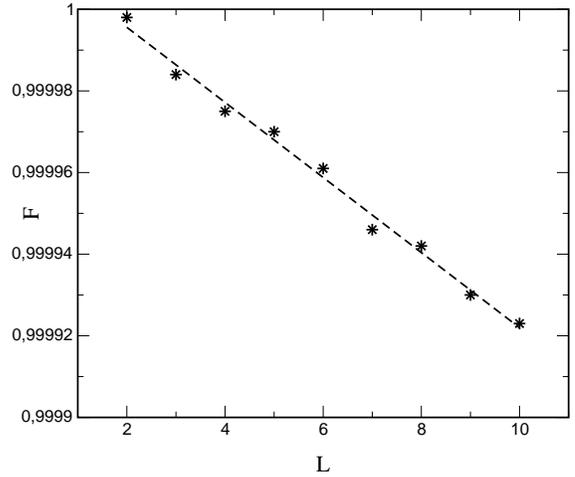}
\caption{
The dynamical fidelity as a function of the number $L$ of spins,
for $J=1.945$ and  $\Omega_k= 0.1216$, $a=100$. Also shown is the
best linear fit with the slope $9.2\pm 0.3 \times 10^{-6}$. }
\label{res}
\end{figure}

Choosing $\Omega$ values as given by Eq.(\ref{tupi}), one gets
that the probability for near-resonant transition is zero,
$\epsilon=0$. So, only non-resonant transitions lead to unwanted
states. In Fig.(\ref{res}) we show the fidelity as a function of
the number of spins $L$ for $\Omega_{k} = 0.1216$. These data
should be compared with the analogous ones indicated by triangles
in Fig.(\ref{fidn}).

\vspace{0.5cm}

As one can see, despite the closeness of these two $\Omega$ values
(less than 3\% of difference), the fidelity is better by two order
of magnitude (see different scales on the $y$-axis). It is clear
that such preferred $\Omega$ values should be chosen in any
practical implementation of the algorithm. However, due to a high
instability of such resonant values, see Fig.(\ref{fidom}), the
detailed analysis can only be done within a more general study
under the presence of small variations in parameters such as
$\Omega, J, a$. This study is currently in progress.

\section{Conclusions}

We have studied the model of a quantum computer that consists of a
one-dimensional chain of 1/2-spins (qubits), placed in a
time-dependent electromagnetic field. The latter is given by a
sequence of {\it rf}-pulses, corresponding to a chosen quantum
protocol that allows to generate an entangled state for remote
qubits from the initial ground state. Main attention is paid to
the analysis of the dynamical fidelity, defined as an overlap of
the actual finite state, with the ideal one determined by the
quantum protocol.

We considered the region of the selective excitation where the
resonant excitations of specific qubits can be implemented by
time-dependent pulses. Analytical treatment of the stationary
Hamiltonian which describes the evolution of the system during a
single pulse, has revealed that in the selective regime the
quantum chaos can not appear. Moreover, in this regime a
perturbation theory can be applied to all quantities of interest.

Our detailed study of the dynamical fidelity manifests excellent
agreement between numerical data and the predictions obtained in
the perturbative approach. In particular, we have found how to
choose parameters of the model in order to get the best dynamical
fidelity for the creation of the remote entangled state. Specific
attention has been paid to the dependence of the dynamical
fidelity on the number $L$ of qubits. We show, both analytically
and numerically, that the dynamical fidelity decreases linearly
with an increase of $L$, and we give an analytical estimate for
the slope of this dependence.

\section{Acknowledgments}

The work of GPB was supported by the Department of Energy (DOE)
under the contract W-7405-ENG-36, by the National Security Agency
(NSA) and Advanced Research and Development Activity (ARDA). FMI
acknowledges the support by CONACyT (Mexico) Grant No. 34668-E. We
acknowledge R.Bonifacio for useful discussions. Authors also
thanks the International Center at Cuernavaca for financial
support during the workshop "Chaos in few and many-body problems"
where this work was started.

\end{document}